# Ensemble Forecasts of Solar Wind Connectivity to 1 $R_s$ using ADAPT-WSA

**D.E. da Silva[1,2,3], S. Wallace[1], C.N. Arge[1], S. Jones[1,5]**

[1] Solar Physics Laboratory, NASA Goddard Spaceflight Center, Greenbelt, MD, USA

[2] Laboratory for Atmospheric and Space Physics, University of Colorado, Boulder, Boulder, CO, USA

[4] Goddard Planetary Heliophysics Institute, University of Maryland, Baltimore County, Baltimore, MD, USA

[5] Catholic University of America, Washington, DC, USA

**Abstract**

The solar wind which arrives at any location in the solar system is, in principle, relatable to the outflow of solar plasma from a single source location. This source location, itself usually being part of a larger coronal hole, is traceable to 1 $R_S$ along the Sun's magnetic field, in which the entire path from 1 $R_S$ to a location in the heliosphere is referred to as the solar wind connectivity. While not directly measurable, the connectivity between the near-Earth solar wind is of particular importance to space weather. The solar wind solar source region can be obtained by leveraging near-sun magnetic field models and a model of the interplanetary solar wind. In this article we present a method for making an ensemble forecast of the connectivity presented as a probability distribution obtained from a weighted collection of individual forecasts from the combined Air Force Data Assimilative Photospheric Flux Transport – Wang Sheeley Arge (ADAPT-WSA) model. The ADAPT model derives the photospheric magnetic field from synchronic magnetogram data, using flux transport physics and ongoing data assimilation processes. The WSA model uses a coupled set of potential field type models to derive the coronal magnetic field, and an empirical relationship to derive the terminal solar wind speed observed at Earth. Our method produces an arbitrary 2D probability distribution capable of reflecting complex source configurations with minimal assumptions about the distribution structure, prepared in a computationally efficient manner.

**Plain Language Summary**

Solar wind which arrives at Earth mostly comes from one place on the sun at a time. There is no measurement of where that place is, but scientific modeling can give us an estimate. In this article, we present one way of making that kind of estimate. Our way gives a probability distribution of where the solar wind is coming from, instead a single estimate. The primary engine of our method is the ADAPT-WSA model. Applications of knowing where the solar wind is coming from on the Sun include helping other studies and deciding where to point solar imagers.

**Key Points**

1. An algorithm for connecting a location in the solar system to its source location using ADAPT-WSA is presented
2. Multiple variations of ADAPT-WSA are collected and summarized into probability distributions using Kernel Density Estimation
3. Usage of source location modeling to inform scientific research and spaceflight mission operations are reviewed and discussed

## Introduction

The solar corona is comprised of complex time-dependent magnetic fields frozen into a highly ionized plasma. The solar magnetic field in this region can instantaneously be categorized into closed field lines, extending no further than a critical point and open field lines, extending far beyond the critical point out to the heliopause. Nevertheless, on long timescales the corona is well approximated by potential fields in which field lines are either open or closed (Riley et al., 2006).

Concentrated regions of open and unipolar magnetic field at the Sun's surface correlate with dark regions observed on the Sun at extreme ultraviolet (EUV) wavelengths, defined as coronal holes. Coronal holes are the primary sources of coronal plasma outflow that forms the solar wind. Models can be used to trace the solar wind from any point in interplanetary space to the corresponding source regions back at the Sun.  Except for limited scenarios involving stream interactions and inter-stream mixing, an observation of solar wind using satellite instrumentation can, in principle, be readily linked to a single source location. The solar wind connectivity is defined for the purposes of this work as the path a parcel of plasma takes from the source location at $1\ R_S$ to a specified travelling satellite. Examples of satellites include Earth and planetary bodies, as well as spacecraft with arbitrary orbits.  The properties of the solar wind observed in situ, such as number density ($n_i$),  proton $T_i$ and electron $T_e$ temperature, radial solar wind speed ($v_{sw}$), and compositional data from mass spectroscopy, are correlated to the dynamics and physical processes acting on the plasma along the path of connectivity, from the source location to the location in the corona where the solar wind parcel fully disconnects from the Sun and flows radially outward.  (e.g. Zurbuchen 2002, Fu et al., 2008; Borovsky et al., 2016; Zhao et al. 2017).

Space weather research and spaceflight operations are aided by modeling efforts which connect the solar wind observed in situ by spacecraft or at an interplanetary body to its specific source region in the low corona. This modeling task is the focus of this article. For example, studies utilizing multi-point spacecraft observations of the solar wind at the scale of the solar system are aided by connecting each observation to a coronal hole, potentially generating a global solar outflow perspective (Gómez-Herrero et al., 2011; Zhu et al., 2018; Dresing et al. 2012). Other research studies analyzing compositional changes in the solar wind are aided by relating to multi-spectral imagery of the source region (Liewer et al. 2005). In such a case, the multi-spectral imagery may provide supportive information regarding per-species abundances and photon-releasing chemical reactions at the source location. In another example, spaceflight missions operating narrow-view solar imagers for solar wind studies must decide where to point their instruments, and often the most useful place to point is the coronal hole linked to an observing spacecraft or body such as the Earth (e.g., for space weather applications) (Rouillard et al., 2020).

In this article, we address the modeling task using the combined ADAPT-WSA model. The output of ADAPT, which are global photospheric magnetic field maps, is used as an input to WSA. WSA, the Wang-Sheeley-Arge model, is a semi-empirical model which predicts the speed and interplanetary magnetic field polarity of the solar wind speed (Arge and Pizzo 2000; Arge et al., 2003a, Arge et al., 2003b; McGregor et al., 2008) (Sheeley et al., 2017; Pizzo et al., 2011; Arge et al., 2000).  ADAPT, the Air Force Data Assimilative Photospheric Flux Transport model (Arge 2010, Arge 2011, 2013), is a model which generates estimates of the global photospheric maps at any specified point in time by evolving the global photospheric magnetic field using flux transport modeling and performing on-going data assimilation of near-side magnetogram measurements (Worden & Harvey 2000; Schrijver and de Rosa 2003; Hickmann et al, 2015). The ADAPT model can assimilate this class of measurements, in this work originating from the Global Oscillation Network Group (GONG) instrument network (Harvey et al., 1996, Plowman et al., 2020). ADAPT itself is an ensemble model and produces a range of global photospheric magnetic field maps (or realizations) that represent different samplings of convection cell parameterizations.

In this work we use a varied set of the velocity and magnetic field output from ADAPT-WSA to create an ensemble prediction of the source location of the in situ observed solar wind. The ensemble dimension is discussed in the methodology section. The result is a probability map of where the source location is most likely to be for a given plasma parcel in the heliosphere. This probability map approach is more powerful than a single point estimate as it inherently highlights uncertainty and consensus in the model variations. The probability map used is flexible in its shape and structure, allowing it to take on non-gaussian forms including bimodal and tailed shapes. For instance, the shape of the probability distribution relative to the edge of coronal hole can speak to the probability distribution of solar wind characteristics, such as $v_{sw}$ (the solar wind speed), in a way a single point estimate cannot.

Previous work on source location estimation has incorporated ensemble methodology to respond to the issue of uncertainty. In the work of Badman et al., 2023, the prediction of solar wind sources for the Parker Solar Probe at 13.3 $R_S$ was studied. Their methodology uses an ensemble set built from a diverse array of models, model parameters, as well physical boundary conditions (Badman et al., 2023). In the work of Koukras et al., 2022, the problem is studied for fast solar wind in particular, where ballistic backmapping is used with PFSS models, and an ensemble set is built from varied solar wind speeds, random noise added to the magnetogram, and varied model boundary locations. The work of Rouillard et al., 2020 studies a related problem of magnetic connectivity in the context of Solar Orbiter, and uses an ensemble set which perturbs an intermediate position in the backmapping algorithm.

The methodology is split into several sections. In the *Connectivity Algorithms* section, we describe the methods used on a single set of WSA velocity/magnetic fields to connect a position in space to a source location at 1 $R_s$. In the *Ensemble Modeling* section, we analyze the theory of the uncertainties involved, how that informs the construction of a representative ensemble set, and how the connectivity algorithm applied to each member of the ensemble set is aggregated into a smooth probability distribution. The *Validation* section presents the results of studies designed to gauge the effectiveness of the model through analysis of secondary characteristics. Finally, the *Conclusion* section summarizes the method presented and reviews applications.

## Connectivity Algorithms

In this section, we describe the algorithm for determining connectivity between a location in the solar system and 1 $R_s$. The methodology of this individual section is identical to that which exists in WSA v5.2, as deployed at the Community Coordinate Modeling Center (CCMC).

The connectivity computed is driven by a single global coronal magnetic field solution over the WSA domain (1 $R_s \leq r < 5.0\ R_s$). This method is repeated for each of the magnetic field solutions in the varied set generated by the different ADAPT map realizations. This algorithm associates a traveling point in the solar system, discussed here as the satellite, with a source location in heliographic latitude and longitude at 1 $R_s$.

WSA is comprised of two potential field (PF) type models. The inner model is a traditional potential field source surface model (Schatten et al., 1989; Altschuler and Newkirk 1969; Wang and Sheeley 1992) with inner boundary at the 1 $R_s$, the photosphere, and source surface at 2.51 $R_s$. Every magnetic field line that reaches the source surface is open by construction, and this boundary effectively serves as the critical point. The Schatten current sheet (SCS) (Schatten et al, 1971) model is used to specify the magnetic field of the outer coronal region and has an inner boundary at 2.49Rs and outer boundary is located at infinity, however the model solution is normally used only out to 5Rs, though different outer boundary radii can be used. A small overlap region between the PFSS and SCS is used in the model coupling. The PFSS magnetic field values at 2.49Rs are used as the inner boundary to the SCS. Because the WSA outer boundary is outside the source surface, and we make the assumption that outside the outer boundary the solar velocity can be approximated by $\vec{v}_{sw} = v_r\,\hat{r}$ while inside the model domain the solar wind must be considered in theory as a three-component vector $\vec{v}_{sw} = v_r\,\hat{r} + v_\theta\,\hat{\theta} + v_\phi\,\hat{\phi}$ (Schatten et al., 1969). The three-component vector $v_{sw}$ within the WSA domain is supplied directly by the magnetic field solution. The plasma beta in the corona is low (magnetically dominated) and therefore the plasma flows along magnetic field lines without strongly perturbing them. Within the WSA coronal domain, there are therefore two sub-domains separate by an interface radius (2.49 $R_s$), known as the PFSS and SCS domains. For more information on these sub-domains, the reader is pointed towards Mayer et al. 2012, McGregor et al., 2008, and McGregor et al., 2005.

The approach we will use is to break the path determining the magnetic field connecting the satellite to the source location into three sub-paths: from the satellite to the model outer boundary ($P_{outer}$), from the outer boundary to the interface radius ($P_{SCS}$), and from the interface radius to the solar surface ($P_{PFSS}$). In this decomposition, each starts where the previous one ends.

**Equation 1: Decomposing the Full Connectivity Path into Three Sub-paths**

$$P_{connectivity} = P_{outer} \cup P_{SCS} \cup P_{PFSS}$$

In *Figure 1* an illustration of $P_{SCS}$ and $P_{PFSS}$ is given in the WSA domain. In this plot, the blue line is the model outer boundary, and the red line the interface radius. In this plot, field lines are generated from the satellite position projected onto the outer boundary (henceforth known as "subsatellite points") over one Carrington rotation. This plot shows how all paths end up in one of several coronal holes, and the field becomes increasingly more radial in the SCS sub-domain until the radial assumption is fully satisfied at the outer boundary. We note that the red and blue lines are drawn on the ecliptic, where-as the subsatellite points are slightly off the ecliptic, which may make it appear that the connection do not continue out to the outer boundary when they in fact do.

# Near-Sun WSA Solar Wind Connectivity (<5 Rs)

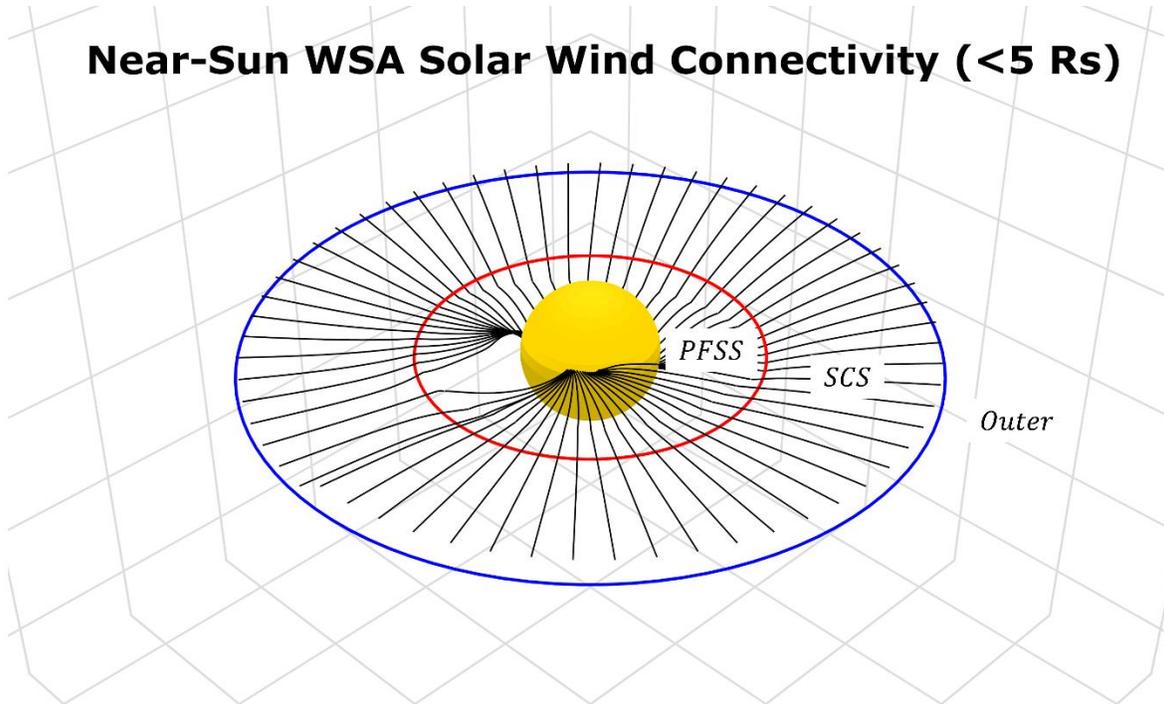

*Figure 1 - Connections between the subsatellite points for Earth at the outer boundary (blue) to the surface of the sun, showing that subsatellite points may be connected to one of several coronal holes. These connections were obtained by tracing the magnetic fields generated by WSA starting at each subsatellite point at the outer boundary (5 $R_s$). The red line represents the interface boundary (2.5 $R_s$) between the Schatten Current Sheet (SCS) sub-domain and the Potential Field Source Surface (PFSS) sub-domain within WSA.*

To solve for $P_{outer}$, we start by utilizing ephemeris data of the satellite. The ephemeris data holding the satellite's position is projected down to the outer boundary of the WSA model (i.e., the subsatellite points are identified). This is done up to a Carrington rotation into the future for the satellite's position at the grid resolution of the model (in this case 2°). This produces a ring on the outer boundary of subsatellite points, and 2nd and 3rd versions of this ring are created by perturbing the latitude by half the grid resolution (in this case $\pm 1°$).

WSA uses an empirical relationship to estimate the solar wind speed at the model outer boundary. The current variation of the equation is given by,

**Equation 2: WSA Empirical Velocity Relationship**

$$v_{sw}(f_s, \theta_b) = 285 + \frac{625}{(1+f_s)^{0.22222}}\left\{1.0 - 0.8e^{-\left(\frac{\theta_b}{2}\right)^2}\right\}^3 \quad \text{km/s}$$

where $f_s$ is the magnetic fied expansion factor and $\theta_b$ is the minimum angular distance that an open field footpoint from the nearest coronal hole boundary (Arge et al 2004, Riley et al., 2015; Wallace et al., 2020).

Radially outflowing point-parcels starting at the outer boundary are then simulated as originating from these three rings using velocity $\vec{v}_{sw}^i = v_r^i \hat{r}$ where $v_r^i$ is obtained from the WSA empirical velocity relationship. These radially outflowing point-parcels are then propagated out to the radius of the satellite position. . The propagation is ballistic except for a simple ad-hoc methodology to prevent fast particles from overtaking slow ones, which results in stream interaction regions reminiscent of the real solar wind (Arge et al., 2000). Stream rarefactions are also produced. This simple point-parcel simulation is extremely fast due to its simplicity and provides a good-enough solution while avoiding overhead (from both operational and computational perspectives) from coupling with an MHD model such as ENLIL.

The output of the point-parcel simulation is $P_{outer}$, providing the connectivity of the satellite with the set of substatellite points at the outer boundary, which are deemed as connected to the satellite at an associated time in the future. The next steps are to find $P_{SCS}$ and $P_{PFSS}$. The sub-path for $P_{SCS}$ is found by tracing the magnetic field line inward from the outer boundary starting at each substatellite point to the interface radius (across the SCS sub-domain), and then the sub-path for $P_{PFSS}$ is found by tracing the field line inward from the interface radius to the model inner boundary, which is the solar surface. Because the solar wind velocity and the magnetic field are aligned (parallel or anti-parallel) in this region by the plasma $\beta$, tracing the magnetic field inward is sufficient for determining the solar wind trajectory. The field line tracing methodology used is an implementation of Runge-Kutta 45 (RK45), where the field at each step is determined by calculating the magnetic field, using spherical harmonics, at the four grid cell centers (at fixed R) bounding the end point position of the last step inward and then interpolating these values to the required position in latitude and longitude. To perform the field line trace, an iterative algorithm steps along the field line in the direction of $a\hat{B}$, solving the equation $\frac{d\mathbf{r}}{dt} = a\hat{B}$ where $\mathbf{r}(t)$ is the position along the trace and $a$ is either $+1$ or $-1$. The value of $a$ is chosen before iteration begins and is selected based on the polarity of the field at the outer boundary, specifically to force the trace move towards the sun even if $\hat{B}$ is pointed outwards.

Supplementary visualizations for understanding of the point-parcel simulation for $P_{outer}$ and the magnetic field tracing for $P_{SCS}$ and $P_{PFSS}$ are found in *Figure 2*. The top panel shows links between points on the first subsatellite ring and the end points on the solar surface residing in coronal holes (colored by their asymptotic final speed per the model). The many points along the ring correspond to very different solar speeds (from about $350 - 700$ km/s). This justifies the need for the point-parcel simulation, since the transit time to, for example Earth, accordingly, varies by a factor of 2X and the corresponding outer boundary location is not obvious.

The bottom panel of *Figure 2* shows a map of the solar wind velocities over the outer boundary sphere determined by using the WSA empirical relationship. This demonstrates the complex longitude-dependence of $v_{sw}$ on the outer boundary, which is most variable near the equator and least variable at high latitudes (Riley et al., 2001; Riley et al., 2015). At this stage, the major uncertainties lie in the ability to estimate the solar wind speed at the model outer boundary (performed through the WSA empirical relationship), and the ability to model parcel's transit between the model outer boundary and the satellite position (performed through the point-parcel simulation).

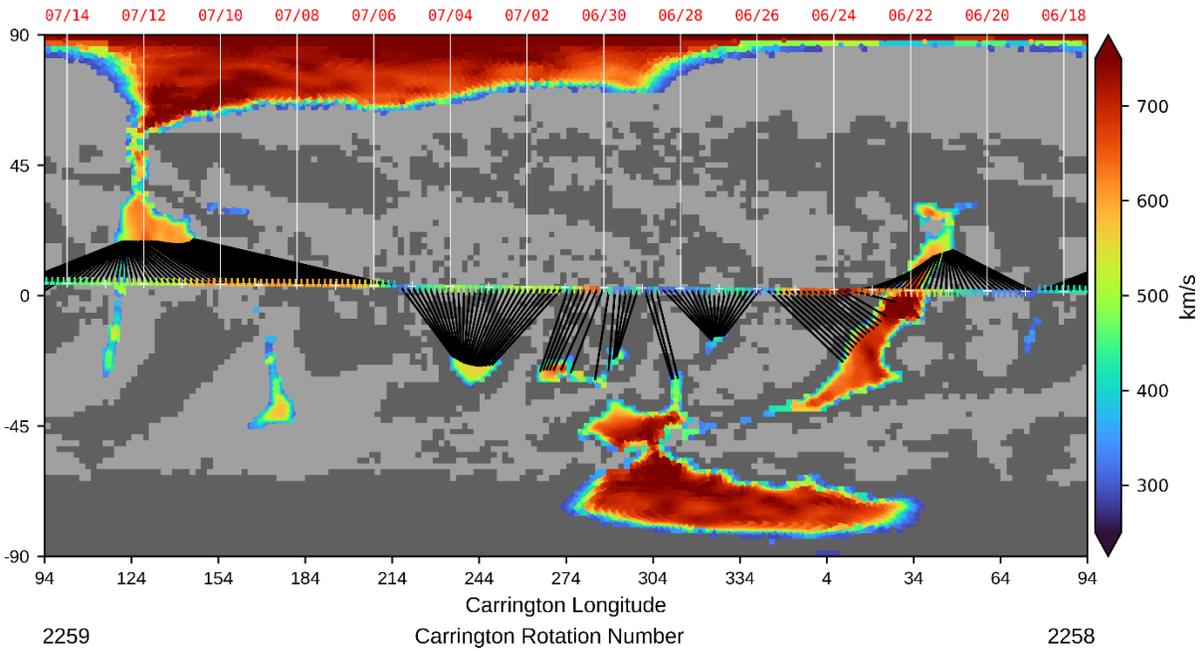

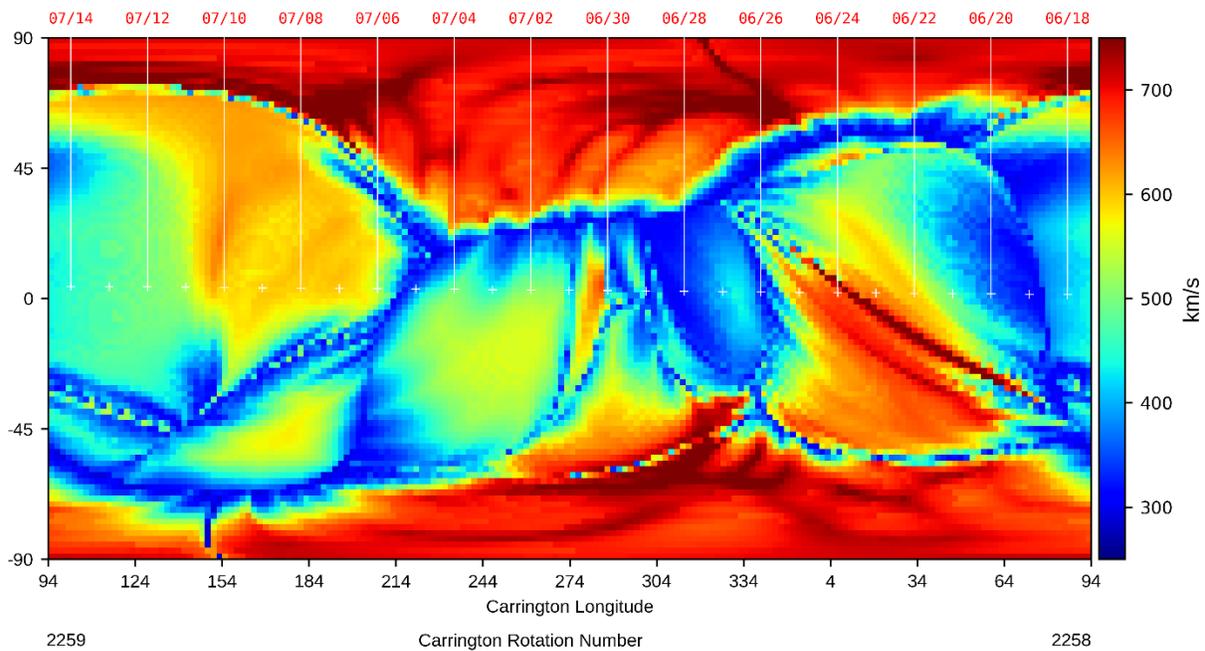

*Figure 2 - Top Panel: WSA-derived coronal holes and subsatellite point connectivity. The black lines show the connectivity between subsatellite points (for Earth) at the model outer boundary (5 $R_s$) and their source location within a coronal hole. The coronal holes are colored by the outflow velocity at each location; with the fastest solar wind originating farthest from the edge. Bottom panel: the solar wind velocity estimated by WSA at model outer boundary. We see a complex velocity field constructed from the combined coronal hole sources, where-in each patch in a coronal hole becomes mapped to a much larger region on the outer boundary surface.*

# Ensemble Modeling

The previous section describes how the connectivity endpoints are determined for individual magnetic field solutions. In this section, we discuss how we turn a collection of forecasts of connectivity endpoints into a probabilistic ensemble forecast for the connectivity endpoint as a random variable, represented as a probability density function (PDF) over space. While past works have made use of ensemble sets to represent uncertainty (Badman et al. 2023; Macneil et al., 2022), our methodology differs in the final presentation PDF using kernel density estimation.

Using the connectivity algorithm, we can provide a collection of connectivity solar surface endpoints $\vec{x_i}$ corresponding to times $t_i$ within a given time window to be discussed below. To create a broad representative collection of connectivity endpoints, which we call the varied set, we utilize multiple solutions originating from the following sources (*Table 1*). Overall, the varied set has about 250-350 members.

**Table 1: Sources of Variation in Ensemble Varied Set**

| Variable | Name | Description | Multiplier |
|---|---|---|---|
| $N_{realizations}$ | ADAPT ensemble variation | Number of ADAPT realizations | 12 (as of publication) |
| $\theta_{subsat}$, $\theta_{subsat} \pm \frac{\Delta\theta}{2}$ | Perturbed physics | The subsatellite points on the model outer boundary used in the point-parcel simulation is perturbed by plus or minus half grid cell, equal to 1°. (see Connectivity Algorithms section) | 3 |
| $\Delta t_{window}$ | Window of acceptable parcel arrivals times around target time | Because solar wind results often show structure delayed or too soon, accept parcel intersections occurring within ±12 hours | Not consistent due to varying $v_{sw}$, but on average around 7 |

The ADAPT model provides a natural source of variation because it provides an ensemble of global photospheric field maps. The ensemble is generated by varying one or more parameters in model. The parameter generally varied in ADAPT corresponds to unobservable supergranular convection cells, which affect the magnetic flux transport, on the far-side and high-latitude regions outside the area of assimilated inputs (Hickmann et al 2015). As of publication, the ADAPT model provides a set of 12 maps (or realizations) and corresponding to an increase of the size of the varied set by a factor of 12.

Furthermore, as mentioned in the Connectivity Algorithms section, we perturb the subsatellite points on the model outer boundary used as starting points for the point-parcel simulation. The perturbation is simply plus or minus half a grid cell, which comes out to 1° at the current grid resolution., which is 2° This increases the size of the varied set by 3X.

Finally, we consider parcels arriving at the satellite within $\pm 12$ hours around the target time. This is justified because solar wind models have been identified in practice to produce correct structures (such as current sheet crossings or high-speed streams) but often offset in time with what is observed at $L_1$. Therefore, considering connections from parcels arriving slightly in the past and the future gives the opportunity to catch events that arrive too soon or that are delayed. Because variation exists in the solar wind speed, parcels will arrive at an unsteady rate, but at model resolution of 2°, about ~7 parcels arrive at earth in a 24-hour interval, which increases the size of the varied set by the same factor.

The next step is to transform our discrete set of point-estimates into a smooth probability distribution for ease of visualization and interpretation. The method used is the Kernel Density Estimate with a Gaussian Kernel (Hastie et al., 2009). This method is simply akin to placing a gaussian distribution at the location of each point estimate, and then adding together all the gaussians to form a final distribution. An example of this in 1 dimension is shown in *Figure 3*. Conceptually, where there are many point-estimates close by, many gaussians will super-impose together to create a high-point of the probability distribution, and when few point-estimates lie within a region, the probability distribution will lack gaussian contributions and remain low.

This has advantages over a histogram in that no "pixelization" effect occurs, as the probability distribution is smooth between points. Compared to fitting a single shape to the data, the kernel density estimate can take on multiple shapes, from bimodal, to tailed and curved, to adapt to the data. This method has been previously applied to ensemble forecasts in the context of weather simulation (Wang and Bishop, 2005; Wilks 2011). While previous work has predicted source location probability distributions using prescribed shapes such as Gaussian Mixture Clustering (Koukras et al., 2022), this is the first work which predicts a source location probability distribution without any a-priori assumptions about its structure.

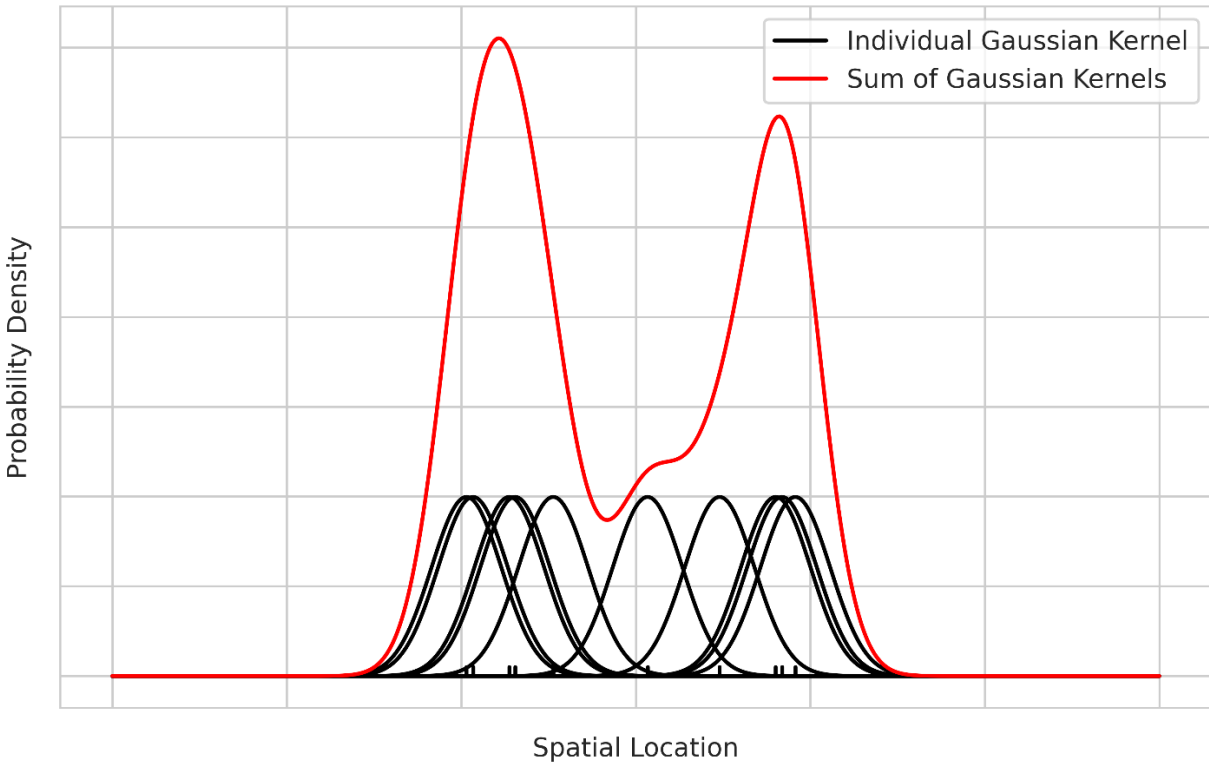

*Figure 3 - 1D Illustration of how the Kernel Density Estimate (KDE) algorithm works to turn a collection of point estimates into a smooth probability distribution. In the method, a gaussian is placed at each of the point estimates, and final smooth probability distribution is the sum of the gaussians. By adding these gaussians, the result will be taller where many points are clustered together and shorter where fewer points occur, with little constraint on its shape. The standard deviation of the gaussians is a parameter which is calibrated in the appendix section Calibration of the Gaussian Kernel Width.*

The equation for the kernel density estimate in two dimensions is given in *Equation 2*. We note that we also utilize a weighting scheme for the gaussians, so-in that some gaussian contribute more than others based on our confidence in that point-estimate. The weighting scheme used here is a "time-decay" weighting scheme based on the $\Delta t_{window}$ component of the varied set. That is, if we are making a probability map for a target time, solar wind parcels in the acceptable window are weighted in an exponentially decreasing fashion based on their time-difference from the target time.

**Equation 3: Kernel Density Estimate and Time-Decay Weighting**

$$f(x) = \frac{1}{\sum_{k=1}^{N} w_i} \sum_{k=1}^{N} w_i G(d(x, x_i); \sigma)$$

$$w_i = e^{-|t_i - t_{target}|/T}$$

In Equation 3, $f(x)$ is the value of the PDF at the location $x$ on the surface of the Sun, nominally given by $(\theta, \phi)$. The values $w_i$ are the weights of each Gaussian. The function $d(x, x_i)$ is the great-circle distance (haversine distance) metric between the surface locations $x$ and $x_i$, in units of degrees. The function $G(\cdot\,; \sigma)$ is the Gaussian PDF with standard deviation $\sigma$. For more information about how the $\sigma$ parameter is calibrated, see the appendix section *Calibration of the Gaussian Kernel Width*. In short, $\sigma$ is chosen to compromise between over and under-smoothing via a maximum likelihood test for the full ensemble. In the definition for the weights $w_i$, the variable $t_i$ is the corresponding time of parcel arrival, $t_{target}$ is the target time for the forecast, and $T$ is a scale time (comparable to a scale height) which defines the rate of exponential decay. A scale time of 3 hours was used so that when $t_i$ is 12 hours away from $t_{target}$, the weight drops to 2% of what it would be if $t_i = t_{target}$.

An example ensemble forecast for the solar wind connectivity source locations at Earth in the form of a PDF can be found in *Figure 4*. The PDF is displayed red/orange/yellow normalized to the peak (red), where it fades to transparent when the value is very close to zero. The background of this map is the photospheric magnetic field Br from a single ADAPT ensemble member. Since the ensemble set of maps generally do not appear significantly different except at the finer scales, selecting any map for display purposes if sufficient. The green regions identify the coronal holes based on coronal predictions by the WSA model. The WSA model is run for each ADAPT map used in the 24-time interval, and they are then all weighted together and normalized (i.e., 1 implies 100% agreement that a pixel is within a coronal hole, while 0 indicates that none of the solutions predicts the pixel resides within a coronal hole). The transparency level is set by the consensus across the varied set that a coronal hole exists at that location.

The PDF shows a bimodal and tailed shape, collectively stating that the connectivity is expected to come from one of two nearby (but distinct) low-latitude coronal holes. For the more western coronal hole, the PDF describes the source location most likely being in the eastern edge of that hole, and for the more eastern hole, spread somewhere across the length (as a tailed distribution). The nominal boundaries of each PDF region (teal lines) are found by searching for closed loops of isolines of a particular very low PDF value (taken here as $10^{-7}$). Once a boundary is found for a region, the PDF can be integrated within the closed loop to summarize the total enclosed probability within that region (yellow text). In this example, we see that each coronal hole is represented about equally in the varied set (51% vs 49%).

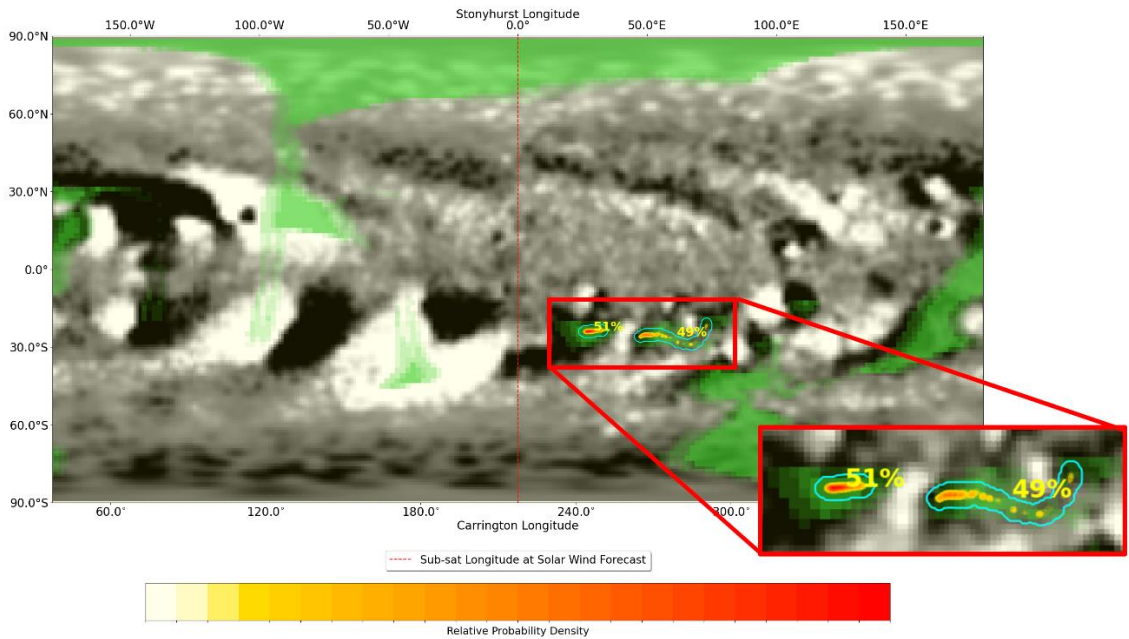

*Figure 4 - An ensemble forecast of the solar wind connectivity at Earth using the method presented in this article. This utilizes the conversion of a set of point estimates of the solar wind source location ($N = 309$) to a smooth probability distribution using Kernel Density Estimation (KDE) with a gaussian kernel. The background coronal holes are colored in green and averaged between the varied set, with the opacity proportional to the consensus that a coronal hole exists at that location. The background magnetogram is taken from a single realization for reference. The probability distribution is normalized to the peak, with red representing the most probable location and yellow representing declining likelihood.*

Uncertainty and LimitationsThere is a fair amount of uncertainty and limitations in virtually all coronal and solar wind models (MacNeice et al., 2009a; MacNeice et al., 2009b; MacNeice et al., 2011; MacNeice et al., 2018; Norquist and Meeks 2010; Norquist et al., 2013; Owens et al., 2005; Owens et al., 2008), due to the lack of knowledge of the specific mechanism(s) heating the corona, as well as those generating and accelerating the solar wind. As a result, both potential-field type models such as WSA, and more advanced magnetohydrodynamic (MHD) codes account for these processes by relying on free parameters that are tuned empirically. Potential field based models are magnetostatic, and as such, the coronal field is derived based on the assumption that the corona is nearly potential. The coronal field is then forced (often prematurely) to be open and radial beyond a certain height (typically 2.5 $R_s$), defined as the source surface. When the source surface (an artificial construct) is raised, PFSS models produce smaller coronal holes and less total open area, while the opposite holds true when this height is lowered. To improve on the original PFSS model, WSA additionally derives the outer coronal field with the SCS model, extending the coronal field to anywhere from 5 – 21.5 $R_s$ before it becomes radial (Wang & Sheeley 1995; Arge et al. 2004; McGregor et al. 2008), providing a more realistic magnetic field topology of the upper corona.

Several studies suggest that the source surface height is dependent on solar cycle, as well as input magnetogram data (e.g. Lee et al., 2001; Arden et al. 2014; Nikolić et al., 2019). In the results shown here, we do not vary the source surface height, as the traditional value has been shown to produce good agreement between WSA-derived coronal holes and those observed in He 10830 nm and EUV on three Carrington rotation averages (Wallace et al., 2019). Thus, one would only need to adjust the source surface height on a case-by-case basis. When using our tool, we advise that any adjustments made to the source surface height be made to within reasonable thresholds as defined by the current literature, and that any model output be validated comparing the model-derived $v_{sw}$ and IMF with that observed in situ.

On the other hand, MHD models use a time-dependent energy equation to generate the coronal magnetic field, where the amount of heating can be modulated to open and close down flux. While it is necessary to use a time-dependent MHD model for certain applications (*e.g.* identifying constraints for coronal heating processes, modeling transient events), a PFSS-based model is better suited for identifying the source regions of the ambient solar wind observed in situ as described in this paper for at least the three following reasons. First, MHD models are not currently capable of deriving the spacecraft magnetic connectivity to the solar surface in a way that is computationally efficient enough to provide routine predictions of the source locations of the observed solar wind. Thus, PFSS models are preferred for applications such as our tool where we are deriving the spacecraft connectivity to the solar surface. While static PFSS-based models have their limitations, using an ensemble of updated synchronic-like photospheric field maps (such as those provided by ADAPT) to drive WSA both minimizes and helps to quantify the uncertainty in the model solution. Second, while MHD and force free models can more accurately determine the coronal magnetic field configuration in non-potential regions such as active regions, the input boundary conditions in these regions are usually the most suspect (*e.g.* most photospheric field maps make use of the line-of-sight magnetic field measurements along with the radial field assumption, which is known to be a very poor assumption in active regions). It is therefore unclear if using more advanced models actually improves magnetic connectivity determinations over that of the PFSS model, given the large uncertainty in the input boundary conditions. Third, both PFSS and MHD solutions have been shown to produce relatively similar coronal topologies at solar minimum (Riley et al., 2006). This is likely due to the lack of knowledge in the detailed physics driving coronal heating and the acceleration of the solar wind, and the empirical nature of both types of models.

Further, a primary source of uncertainty for all coronal models are the large uncertainties in the global photospheric magnetic field maps, which serve as a key driver to these models (see Posner et al., 2021 for a detailed review of this problem). One major issue is that polar fields are not observed, yet they are a crucial component to the global coronal field (Arge and Pizzo 2000; Riley et al., 2019}. Similarly, global coronal and solar wind solutions can be significantly impacted by the lack of far-side observations (Arge et al., 2013; Cash et al., 2015; Wallace et al., 2022). This can occur when a new active region emerges on the Sun's far-side and therefore is not included in the photospheric field map. Coronal solutions can also be adversely impacted when these new active regions rotate onto the solar near-side and ``suddenly'' appear in the maps. The effects of this can be mitigated in retrospective analyses, where we can allow for new far-side emergence to assimilate into the solution for 1-2 days after a period of interest, which has been shown in previous work to stabilize the coronal solution (Wallace et al., 2022). We also validate our model results with $v_{sw}$ and $B_r$ observations in retrospective studies. While this issue cannot be mitigated for forecasting, our tool helps improve forecasts which suffer the most from the lack of global photospheric field observations, by quantifying the uncertainty in the source location of the solar wind observed at various spacecraft or satellites through a varied ensemble set of estimates.

Lastly, it is important to note that it is not possible for PFSS models to capture the Sun's time dependent phenomena associated with the opening and closing of magnetic flux (*e.g.,* magnetic reconnection, transient eruptions such as coronal mass ejections). While we can account for time-dependent photospheric phenomena with ADAPT, WSA only derives the magnetic connectivity between an observing spacecraft and model-derived field lines that are open. Similarly, WSA cannot provide information regarding how long a particular field line has been open. Therefore, when the model predicts that a spacecraft measured plasma near the time-dependent magnetic open-closed boundary, the two physical scenarios that are possible are 1) the plasma originated from that particular open field line, or 2) the plasma originated on closed field that was recently opened via interchange reconnection, whereas WSA cannot make the distinction between the two possible scenarios. Reliably determining how long the magnetic field has been open at the solar wind source region is currently beyond the capability of *all* global coronal models due to the inaccuracies of the input boundary conditions (*i.e.* photospheric field maps). However, the WSA model can reliably distinguish when the in situ observed solar wind originates from intermittently open field at the magnetic open-closed boundary (i.e. roughly the coronal hole boundary), or from sources that are deep inside coronal holes where the field is continuously open. This information is critical for studies on how the solar wind forms, and for characterizing the in situ properties of solar wind from these two distinct sources.

## Validation

Validating a connectivity forecast is difficult due to the lack of a ground truth dataset (Arge et al, 2023). It is possible to compare to other models, however without the ability to verify those other models against ground truth, the comparison can at best conclude both models in agreement without claim that either is correct. However, it is possible to use secondary features to compare a forecast with observation, not giving a full guarantee but still determining a baseline of validity. Overall, such a comparison can only give modest confidence and provide a sanity-check; it cannot provide conclusive proof that the model is valid. We note that on a case-by-case basis, a number of secondary features can be used to support validation. Such secondary features include the darkness of the source location EUV imagery, whether the observed solar wind is fast/slow, and observed solar wind composition characteristics such as first ionization potential (Young et al., 2018; Young et al., 2021).

The secondary features we will compare are the distribution of the IMF polarity at the photosphere (-1 or +1) provided by our model, and the similar IMF polarity observed at the spacecraft using in-situ instrumentation. Until periods of current sheet crossings, the distribution of IMF polarities should begin dominated by one polarity, then split during the current sheet crossing, and return to being dominated a single polarity. A similar pattern should be visible in the satellite-observed IMF polarity. While one could in principle do a similar comparison between model and observation with the solar wind speed, in practice this is difficult because model solar wind speed forecasts are prone to large errors (i.e., Owens et al., 2008) and one can forecast the correct speed while getting the source location wrong. While consistency between model predictions of speed and IMF polarity with observations provides greater confidence that the model is representing the overall state of the corona and solar wind well, IMF polarity is a much more reliable and definitive indicator as to whether the model is not magnetically connected to the correct region (i.e., if the model predicts the wrong polarity, one knows the model forecast is incorrect).

In Figure 5, we compare the distribution of IMF polarities with data from ACE, STEREO-A, and STEREO-B during the period of March 2010 to August 2010. This provides a multi-point perspective on the model quality for three locations at roughly 1 AU, with each location experiencing current sheet crossings at different times. In this plot, the model line (colored) shows (a) the IMF polarity obtained from the varied sets, averaged over a four-day window, and (b) the IMF polarity observed using in-situ instrumentation on the corresponding spacecraft averaged over a four-day window.

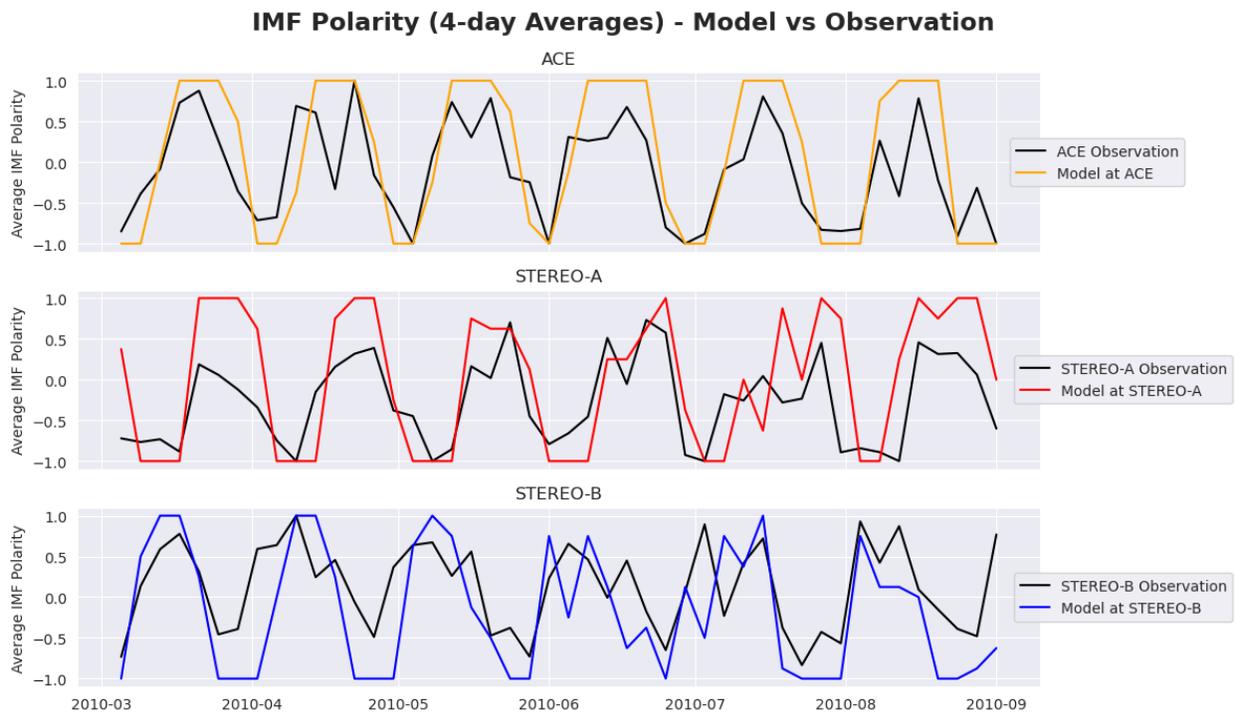

*Figure 5 - Comparison of average of model distribution of IMF polarities versus the average of the observed IMF polarities, for three locations in the solar system with corresponding in-situ observations. The size of the averaging window is four days. In this plot, we notice the broad agreement as to when the current sheet is crossed and the polarity switches. This gives us confidence (though not absolute guarantee) that the model is accurately predicting the correct coronal hole.*

In this plot, we can see that there is broad agreement as to when a current sheet is crossed and when the polarity switches. Between the three locations (ACE, STEREO-A, STEREO-B), the approximate times between polarity switches is approximately half a Carrington rotation for each, though the times of the polarity switches are not in phase due to their differing locations in the solar system. This is largely consistent with the structure of the heliospheric current during this period, which is largely that of titled dipole. This gives us confidence that the model is accurately predicting the correct coronal hole (at least as far as to distinguish between positive and negative polarity coronal holes) and switching its prediction at the correct times. Should the model be picking a coronal hole at random, or in a way which is minimally correlated with the true coronal hole, it would be statistically unlikely to reproduce the level of correlation shown here.

# Conclusion

In this work we present a method for preparing a set of ensemble predictions of solar wind connectivity in the form of a probability distribution. The method uses near-sun magnetic fields obtained from the ADAPT-WSA model and combines those fields with a parcel-propagation method to simulate the arrival of solar wind at a target in the solar system. This produces an ensemble of point-estimates of the solar wind source location, which is then post-processed into a smooth probability distribution using a method known as kernel density estimation.

Estimation of the solar wind connectivity has applications for both scientific investigation and operations. For research applications, tools such as the one presented here are the only way to connect in situ solar wind observations to their specific source at the Sun observed remotely.  Remote sensing of the corona, through multispectral imagery and spectroscopy techniques, reveals both the source type (e.g. active region or quiet sun at the magnetic open-closed boundary,  or deep inside a coronal hole), as well as its associated properties, including composition and temperature. This information is vital to answering open questions in Heliophysics, including how the solar wind is formed, from its origin, to its subsequent release, heating, and acceleration (Viall & Borovsky, 2020).  In operations, solar imagers with narrow field of view must choose where to point their instruments, and having knowledge of the upcoming solar wind source region allows such instruments to point in the most useful direction.

The task of estimating solar wind connectivity is a challenge for the modeling community because there is no ground truth to compare against. The major uncertainties are (a) the ability to estimate the near-sun magnetic field, (b) the ability to accurately estimate the solar wind speed at the outer boundary of the coronal model, and (c) the ability to accurately model the transit of solar wind (including stream interactions) from the outer corona to the target. In this work, we tackle (a)-(b) using the ADAPT-WSA model previously established in the community and present a method for approaching (c) which balances computational efficiency with reasonable detail.

Future avenues for research include assessing expectations of modeled source location uncertainty in relation to key parameters. A central question to answer is: how do measures of uncertainty change in relation to key parameters such as (a) position in the solar cycle, (b) coronal hole longitude/latitude, and (c) physical distance between the final location and the sun? By addressing these questions, we acquire insight regarding fundamental qualities of potential field models and gain physical intuition about the nature of coronal holes themselves. Our method, which is powerful in its ability to represent complex source configurations with minimal assumptions about the probability distribution structure, is uniquely well suited to studying these questions.

# Data Availability Statement

Assimilated GONG measurements in the form of ADAPT maps which describe the radial magnetic field at every point around the sun can be obtained directly for no charge from the National Solar Observatory (NSO) at https://gong.nso.edu/ . Runs of ADAPT-WSA can be performed for no charge through the Community Coordinated Modeling Center's (CCMC's) Run-on-Request system at

[https://ccmc.gsfc.nasa.gov/models/WSA~v.5.2/](https://ccmc.gsfc.nasa.gov/models/WSA~v.5.2/) . Data from the ACE, STEREO-A, and STEREO-B satellites can be obtained from NASA's Space Physics Data Facility at [https://spdf.gsfc.nasa.gov/](https://spdf.gsfc.nasa.gov/).

Python code which implements the methodology of this paper and generates plots similar to Figure 4 can be found open-source on Github at https://github.com/ddasilva/ensemble-solar-wind-connectivity/.

# Appendix

## Calibration of the Gaussian Kernel Width ($\sigma$)

**Equation 3** encodes the model of the kernel density estimation which is used to convert $N$ point estimates of the solar wind source location (where $N$ varies approximately between 250 and 350) into a smooth probability distribution. The method works by summing together $N$ small gaussian distributions, each one centered at one of the point estimates. A free variable in this equation is $\sigma$, the standard deviation used for the gaussian distributions. This variable is calibrated using data, and the method for doing so is the focus of this appendix section. The value of this variable was estimated to be $0.31°$.

The calibration of $\sigma$ can be thought of in terms of in the extremes where $\sigma$ is much too small, and when $\sigma$ is much too large. As $\sigma$ becomes very small, the resultant probability distribution in **Equation 3** approaches a set of Dirac deltas centered on each point in the varied set. When this happens, the distribution loses its ability to interpolate a probability value in between elements of the varied set. In this sense, it is under-smoothed. When $\sigma$ becomes much too large, the opposite happens, and the resultant distribution becomes over-smoothed. If one imagines the case where $\sigma$ is very large (say, 360°), the resultant distribution becomes roughly "flat everywhere" and approaches a uniform distribution. Neither the under-smoothed cased nor the over-smoothed case is desirable, but logically there should exist a middle-ground $\sigma$ which meets the intention of being appropriately-smoothed.

The method used for estimating $\sigma$ is maximum likelihood estimation (MLE) (Hastie et al., 2009). To use MLE, we designed an experiment where we split the varied set for a given map into a training set (90% size) and a test set (10%) size. The logic used is to test multiple candidates for $\sigma$ to find the one which maximizes the likelihood of the resultant distribution predicting the elements of the test set. This likelihood is encoded in **Equation 3** below, where $L(\sigma)$ is the likelihood function to be maximized, Test is the set of test points, and $f(x_i; \sigma)$ is the value of the resultant distribution PDF at the location $x_i$ on the surface of the Sun parameterized by candidate parameter $\sigma$ (see **Equation 2**). The product in **Equation 3** originates from the notion that we are seeking to quantify the probability of finding $x_1$ and $x_2$ and $x_3$ and so on.

**Equation 3: Likelihood Function to Find $\sigma$**

$$L(\sigma) = \prod_{x_i \in \text{Test}} f(x_i; \sigma)$$

To illustrate the behavior of $L(\sigma)$, we will go back to our descriptions of under-smoothed ($\sigma$ too small) and over-smoothed ($\sigma$ too large) distributions, if the resultant distribution is drastically under-smoothed, then it will predict probabilities too low at each point in the test set, because the gaussian has already dropped off. If the distribution is drastically over-smoothed, it will predict too low as well, because each gaussian is normalized to have an area under the curve of 1, and in order to span a large area it must have lower probability density at each point in that area.

In our experiment, we performed this MLE using varied set once every 6 hours for predictions at ACE, STEREO-A, and STEREO-B between January 2010 and March 2010. Each varied set produced a single estimate of $\sigma$, which we collected and analyzed in aggregate. This produced many estimates of $\sigma$, ranging between 0.06° up to 1.00°, with a mean of 0.31° and a standard deviation of 0.23°. The distribution had a positive skew of 1.15 (tail extending to the right). To use a single estimate for operational efficiency, the mean of 0.31° was selected as the outcome of the experiment.